\documentclass[aps,prb,twocolumn,superscriptaddress]{revtex4-1}
\usepackage{amssymb}
\usepackage[utf8]{inputenc}
\usepackage{bm,color,subfigure,amsmath,graphicx}

\DeclareGraphicsExtensions{.pdf,.jpeg,.png}
\providecommand{\sech}{\text{sech}}

\newcommand{\R}{\mathbb{R}}

\newcommand{\hspc[1]}{\hspace{#1pt}}
\newcommand{\beq}{\begin{equation}}
\newcommand{\eeq}{\end{equation}}
\newcommand{\beqp}{\begin{equation*}}
\newcommand{\eeqp}{\end{equation*}}
\newcommand{\bea}{\begin{align}}
\newcommand{\eea}{\end{align}}
\newcommand{\beap}{\begin{align*}}
\newcommand{\eeap}{\end{align*}}
\newcommand{\bxi}{\boldsymbol{\xi}}
\newcommand{\bV}{\mathbf{v}}
\newcommand{\bW}{\boldsymbol{\mathcal{W}}}

\begin{document}

\title{Effects of Thermal Perturbations on Magnetic Dissipative Droplet Solitons}
\author{P.~Wills}
\affiliation{Department of Applied Mathematics, University of Colorado, Boulder, Colorado 80302, USA}
\author{E.~Iacocca}
\affiliation{Department of Applied Mathematics, University of Colorado, Boulder, Colorado 80302, USA}
\affiliation{Department of Applied Physics, Division for condensed matter theory, Chalmers University of Technology, 412 96, Gothenburg, Sweden}
\author{M.~A.~Hoefer}
\affiliation{Department of Applied Mathematics, University of Colorado, Boulder, Colorado 80302, USA}

\begin{abstract}
  The magnetic dissipative droplet is a strongly nonlinear wave
  structure that can be stabilized in a thin film ferromagnet
  exhibiting perpendicular magnetic anisotropy by use of spin transfer
  torque. These structures have been observed experimentally at room
  temperature, showcasing their robustness against noise. Here, we
  quantify the effects of thermal noise by deriving the stochastic
  equations of motion for a droplet based on soliton perturbation
  theory. First, it is found that deterministic droplets
  are linearly unstable at large bias currents, subject to a drift
  instability. When the droplet is linearly stable, our framework
  allows us to analytically compute the droplet's generation linewidth
  and center variance. Additionally, we study the influence of
  non-local and Oersted fields with micromagnetic simulations,
  providing insight into their effect on the generation
  linewidth. These results motivate detailed experiments on the
  current and temperature-dependent linewidth as well as drift
  instability statistics of droplets, which are important
  figures-of-merit in the prospect of droplet-based applications.
\end{abstract}
\maketitle

\section{Introduction}

Localized magnetic textures have recently attracted significant
research interest due to their potential application in logic,
storage, and communication technologies. From the perspective of logic
and storage, static Skyrmions~\cite{Nagaosa2013} are very interesting
textures due to their topological protection against perturbations,
small sizes, and controllable
motion~\cite{Romming2013,Sampaio2013}. On the other hand,
communication applications could benefit from dynamical textures,
notably topological, dynamical Skyrmions~\cite{Zhou2015} and
non-topological, magnetic dissipative
droplets~\cite{Hoefer2010,Hoefer2012b,Maiden2013,Bookman2013,Iacocca2014}.

\begin{figure}
  \begin{center}
    \includegraphics[width=3.3in]{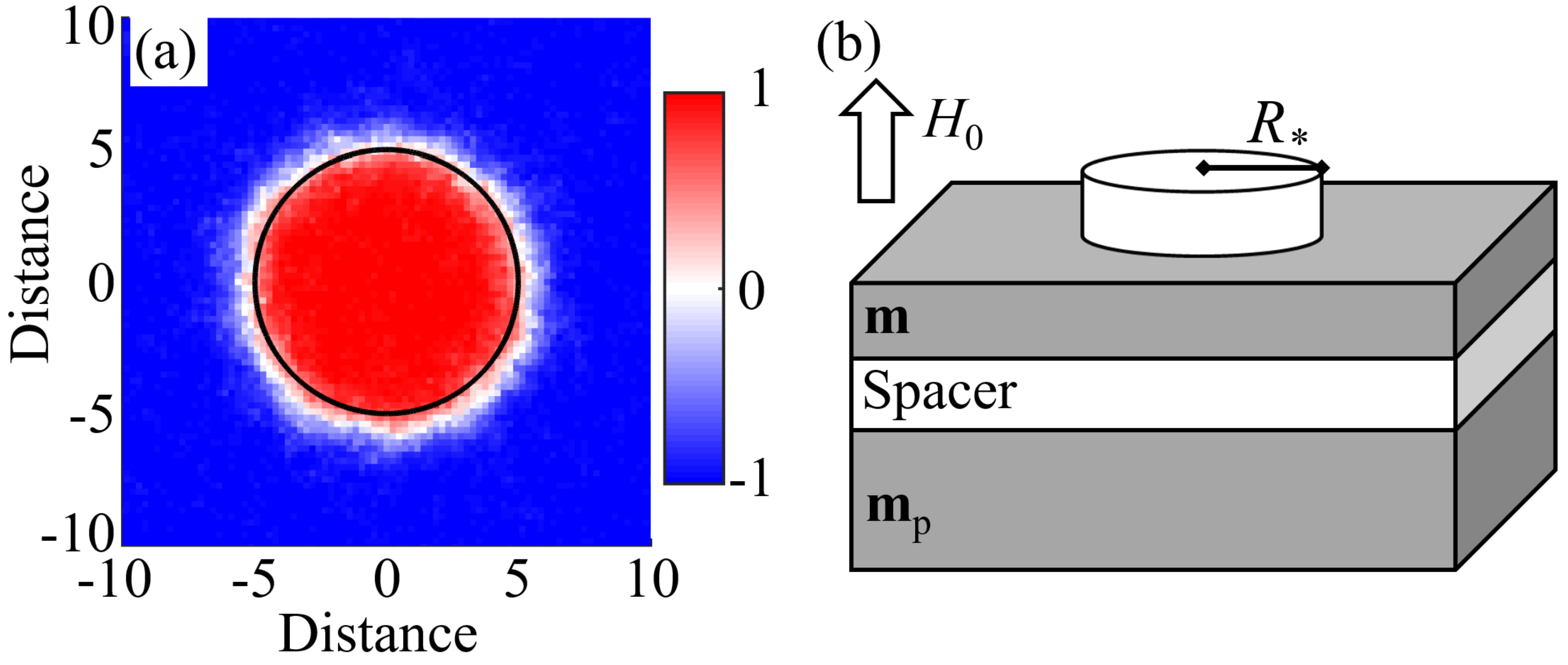}
    \caption{\label{fig:sto} (color online) (a) Typical dissipative
      droplet obtained from micromagnetic simulations at a finite
      temperature. The $\hat{z}$ component of the magnetization is
      quantified by the color scale. (b) Schematic of a NC-STO based
      on a pseudo spin valve trilayer. The free, $\mathbf{m}$, and
      polarizer, $\mathbf{m}_{\rm p}$, magnetic layers are decoupled
      by a non-magnetic spacer. A NC of radius $R_*$ is placed on top
      of the free layer to achieve high current densities. An external
      field $H_0$ is applied perpendicularly to the plane.}
	\end{center}
\end{figure}

Magnetic dissipative droplets (``droplets'' hereafter) have been
widely observed in experiments both at cryogenic~\cite{Macia2014} and
room
temperatures~\cite{Mohseni2013,Backes2015,Lendinez2015,Chung2015}. Droplets
exist in magnetic thin films composed of materials with perpendicular
magnetic anisotropy (PMA)~\cite{Bruno1989}, \emph{i.e.}, in which the
easy axis lies normal to the plane, so that it balances the exchange
energy in favor of a localized
structure~\cite{Kosevich1990,Hoefer2010},
Fig.~\ref{fig:sto}(a). Furthermore, magnetic damping must also be
balanced in order to sustain the droplet in time due to its lack of
topology. To date, this has been achieved by using spin transfer
torque (STT)~\cite{Berger1996,Slonczewski1996} in devices known as
nanocontact spin torque oscillators, NC-STOs~\cite{Dumas2014}. NC-STOs
are composed of a pseudo spin valve where two magnetic layers are
decoupled by a non-magnetic spacer, as shown in the schematic of
Fig.~\ref{fig:sto}(b). The topmost magnetic layer, $\mathbf{m}$, is
where the droplet nucleates and it is usually referred to as the free
layer. The bottom magnetic layer, $\mathbf{m}_{\rm p}$ serves as a
spin-polarizer and it is known as the polarizer or fixed layer. In
order to achieve sufficient current density to oppose magnetic
damping, a nanocontact (NC) of radius $R_*$ is placed on top of the
free layer, confining the current to flow in an approximately
cylindrical path~\cite{Petit2012b} and therefore defining a region of
effectively zero damping in the free layer. An external, perpendicular
applied field $H_0$ is generally used in NC-STOs both to tilt the
polarizer (useful for increasing STT and magnetoresistance), to
provide an external source for the Larmor frequency, and to stabilize
the droplet~\cite{Bookman2013}.

Since the first experimental observation of
droplets~\cite{Mohseni2013}, recent results have investigated
theoretical
predictions~\cite{Hoefer2010,Hoefer2012b,Hoefer2012,Bookman2013,Bookman2015},
shown the existence of hysteresis both at room and cryogenic
temperatures~\cite{Macia2014,Lendinez2015}, identified a well-defined
nucleation boundary~\cite{Chung2015}, and even imaged the droplet via
X-ray magnetic circular dichroism (XMCD)~\cite{Backes2015}. The same
studies have demonstrated the existence of characteristics consistent
with random droplet dynamics, notably low-frequency spectral
features. These have been associated with the droplet exiting the NC
region and succumbing to damping, a drift instability, originating
from the spatial energy landscape created by the current-induced
Oersted field~\cite{Hoefer2010,Petit2012b,Dumas2013,Madami2015} and
externally applied fields~\cite{Bookman2013,Mohseni2013,Bookman2015}
or fluctuations in the material anisotropy spatial
distribution~\cite{Lendinez2015}. However, the relationships between
drift instabilities and physical sources of randomness have not been
established. To provide an analytical understanding of drift
instabilities, we study the effect of thermal noise on droplet
dynamics.

In this paper, we develop the stochastic evolution of droplet dynamics
based on soliton perturbation theory~\cite{Bookman2015} and obtain
statistical observables such as the droplet center variance and the
generation linewidth~\cite{Silva2010}. These results are analytically
obtained by linearizing the equations of motion. From the
linearization, we uncover a deterministic regime of drift instability,
missed by previous analytical works~\cite{Bookman2013,Bookman2015},
where high bias currents induce growth of the droplet velocity on a
long timescale. Randomness can also cause an otherwise
deterministically stable droplet to be expelled from the NC region
when thermal fluctuations are taken into account.  We determine that
such events are extremely rare relative to the precessional timescale
(10-100 picoseconds) but become quite relevant for the typical time
scales of experiments (seconds or more).  Observation, let alone
quantification, of both the deterministic drift instability and the
stochastic rare events is practically unfeasible utilizing standard
deterministic~\cite{Hoefer2010} or stochastic~\cite{Lendinez2015}
micromagnetic simulations alone.  For a stable droplet, the generation
linewidth is found to be dominated by the phase noise induced by a
Wiener process or random walk, linearly proportional to temperature
and inversely proportional to the NC radius. The droplet's center can
be described by an Ornstein-Uhlenbeck (O-U) process with STT acting as
an attractive mechanism that draws the droplet to the center of the
NC.  The determination of both stochastic processes requires subtle
higher order effects from soliton perturbation
theory~\cite{Bookman2013,Bookman2015}.  Full-scale micromagnetic
simulations qualitatively agree with the analytical results, even when
the current-induced Oersted field is taken into account.

The paper is organized as follows. Section II describes the formalism
used to obtain the stochastic equations for droplet dynamics. Section
III explores the deterministic linearization where we obtain the
fundamental droplet dynamical state and linear stability
conditions. Stochastic terms are incorporated into the analysis in
section IV, leading to analytical solutions for the droplet center
variance and generation linewidth at low
  temperatures. Numerical simulations of the nonlinear stochastic
system are presented in section V, demonstrating excellent agreement
with the linearized analytical results. Full-scale micromagnetic
simulations are used to explore regimes of small NC radii, non-local
dipole fields, and Oersted field, beyond the scope of the asymptotic
theory, nevertheless demonstrating qualitative agreement. Finally, we
provide a discussion and concluding remarks in section VI.

\section{Droplet perturbation theory}

The analytical study of droplet dynamics can be approached using
perturbation theory with the magnetic damping and STT coefficients
assumed small. This assumption alone yields droplet nucleation
conditions and the resultant droplet's frequency tunability via
current and field~\cite{Hoefer2010,Chung2015}.  A semi-analytical
generalization can be used to describe coarse droplet motion and
control~\cite{Hoefer2012}.  The additional assumption of a
sufficiently large NC diameter implies a slowly precessing, circular
domain wall description for the droplet~\cite{Bookman2013}, which
enables a detailed analytical description of droplet dynamics in the
presence of physical perturbation~\cite{Bookman2013,Bookman2015}.
This latter regime is the one considered here.

The equation of motion for the free layer magnetization $\mathbf{m}$
is the Landau-Lifshitz equation for a thin, two-dimensional magnetic
film
\beq
\label{eq:LLp}
\frac{\partial \mathbf{m}}{\partial t} = -\mathbf{m}\times
\mathbf{h}_{\rm eff} + \mathbf{p}, \quad \mathbf{m}:\R^2\times\R \to
\mathbb{S}^2 ,
\eeq
expressed here in nondimensional form. The effective field,
\begin{equation}
  \label{eq:1}
  \mathbf{h}_{\rm eff} = h_0 \mathbf{z} + \nabla^2 \mathbf{m} + m_z
  \mathbf{z},
\end{equation}
includes contributions from a perpendicular external field $h_0$, the
exchange field $\nabla^2 \mathbf{m}$, and a perpendicular magnetic
anisotropy (PMA) field sufficient to overcome the thin-film limit of
the demagnetizing field.  Hence the $m_z \mathbf{z}$ term in the
effective field has a positive coefficient, here scaled to unity. This
form of the LL equation, with $\lvert\mathbf{m}\rvert=1$, uses the
time scale $\tau = (|\gamma|\mu_0\eta)^{-1}$, where $\gamma$ is the
gyromagnetic ratio, $\mu_0$ is the vacuum permeability,
$\eta=M_s(Q-1)$ is the field scaling, $M_s$ is the free layer's
saturation magnetization, $Q=H_k/M_s$ is the nondimensionalized form
of the PMA field $H_k$, and the length scale
$L=\lambda_\text{ex}/\sqrt{Q-1}$ where $\lambda_\text{ex}$ is the
exchange length. The NC radius $R_*$ is nondimensionalized to $\rho_*
= R_*/L$. We consider a small perturbation
$\lvert\mathbf{p}\rvert\ll1$ satisfying $\mathbf{p}\cdot\mathbf{m}=0$
in order to preserve constant magnetization magnitude. The
perturbation term considered here includes damping, STT as imposed by
a NC-STO, and a thermal random field~\cite{Brown1963} \beq
\label{eq:p}
\mathbf{p} = \underbrace{-\alpha \mathbf{m}\times(\mathbf{m}\times
  \mathbf{h}_\text{eff})}_\text{damping}+\underbrace{\sigma
  \mathcal{H}(\mathbf{x})
  \mathbf{m}\times\mathbf{m}\times\mathbf{m}_\mathrm{p}}_\text{NC-STO}
- \underbrace{\mathbf{m}\times\mathbf{h}}_\text{thermal}, \eeq where
$0<\alpha\ll1$ is the damping parameter, 
\begin{equation*}
  \mathcal{H}(\mathbf{x}) =
  \begin{cases}
    1 & |\mathbf{x}| \le \rho_* \\
    0 & \mathrm{else}
  \end{cases}
\end{equation*}
is a shifted Heaviside function describing the current path below the
NC,
$\mathbf{m}_\mathrm{p}$ is the normalized polarizer orientation, and
$\sigma=I/I_0$ is the nondimensionalized form of the current $I$,
scaled by
\begin{equation}
	I_0 = \frac{4 \mu_0M_s^2(Q-1) e \pi R_*^2 \delta }{\hbar \epsilon}.
\end{equation}
Here, $e$ is the charge of the electron, $\delta$ is the thickness of
the free layer, $\epsilon$ is the spin torque efficiency, and $\hbar$
is Planck's constant. Example scalings for recent experiments are
listed in Table~\ref{table:scalings} for reference.
\begin{table}[b]
  \begin{ruledtabular}
    \begin{tabular}{ |l|c|c|} 
      Parameters & Refs.~\onlinecite{Macia2014} and \onlinecite{Backes2015}  &
      Ref.~\onlinecite{Mohseni2013}
      \\  
      \hline
      $\tau$ (time, ns) 		&	0.13 	& 0.083		\\
      $L$ (length, nm)			&	13.2 	& 9.25 		\\
      $\eta$ (field, kA/m)		&  198.9 	& 318.1		\\ 
      $I_0$(current, mA)		&  152.75	& 139.7		\\ 
      $T_0$(temperature, kK) 	&  156.0	& 337.2		\\
      \hline
      $\sigma / \alpha$ (scaled current to damping) \,\,\,\,\,\,\,	&	1.96	&	6.44	\\
      $h_0$ (scaled applied field)	&	0.5		&	2.5		\\
      $\rho_*$	(scaled nanocontact radius)	&	5.96	&	5.95
    \end{tabular}
    \caption{\label{table:scalings} Time, length, field, current, and
      temperature scalings and typical nondimensionalized experimental parameters for recent experiments.}
  \end{ruledtabular}
\end{table}

The thermal field $\mathbf{h}(\mathbf{x},t)$ induces random
fluctuations in the magnetization of a small material volume $V$ and
is assumed to be delta-correlated in space and time \emph{i.e.}, white
noise~\cite{Brown1963}. The variance of the nondimensional field is
$\text{Var}[\mathbf{h}(\mathbf{x},t)] = \beta^2$ with
\beq \beta^2=\frac{T}{T_0}, \quad T_0 = \frac{\mu_0 M_s^2 V}{2\alpha
  k_B},
\label{eqn:beta2}
\eeq
where $k_B$ is the Boltzmann constant, $V=\lambda_{ex}^2\delta$ is the
characteristic micromagnetic volume, and $T_0$ is the nondimensional
scaling of the absolute temperature $T$.  Table \ref{table:scalings}
includes typical temperature scalings for recent experiments and our
micromagnetic simulations. The perturbative theory utilized here is
valid in the low temperature regime where $\beta\ll1$. The variance in
Eq.~(\ref{eqn:beta2}) can be dimensionalized by multiplying
(\ref{eqn:beta2}) by $\tau M_s^2 (Q-1)^2$.

The droplet is characterized by its center position
$\boldsymbol{\xi}$, velocity $\mathbf{v}$, collective phase $\phi$,
and precessional frequency $\omega$. In the regime $0\leq v \ll \omega
\ll1$, where $v = |\mathbf{v}|$ is the droplet speed, the droplet
takes on the approximate form of a slowly precessing circular domain
wall with a spatial phase proportional to the droplet's
speed~\cite{Bookman2015}
\begin{subequations}
  \begin{align}
    \cos \Theta &= \tanh \left(\rho-\frac{1}{\omega}\right),\\
    \label{eq:5}
    \Phi &= h_0 t -\frac{\mathbf{v} \cdot
      \hat{\boldsymbol{\rho}}}{\omega^2} + \phi, \quad \phi = \omega
    t + \phi_0.
  \end{align}
  \label{eqn:approxdrop}
\end{subequations}
Equation~(\ref{eqn:approxdrop}) describes the magnetization
orientation of the droplet in spherical coordinates $(\Theta,\Phi)$
with polar angle from vertical $0 \le \Theta < \pi$ and azimuthal
angle $\Phi$.  In Eq.~(\ref{eqn:approxdrop}), we employ
droplet-centered polar coordinates in the plane, so that the radial
unit vector $\hat{\boldsymbol{\rho}}$ points from the droplet center
$\bxi$ to a point in space $\mathbf{x} = (\rho \cos \varphi, \rho \sin
\varphi)$ and the angular unit vector $\hat{\boldsymbol{\varphi}}$ is
orthogonal $\hat{\boldsymbol{\varphi}} \cdot \hat{\boldsymbol{\rho}} =
0$ and satisfies the right hand rule $\hat{\boldsymbol{\rho}} \times
\hat{\boldsymbol{\varphi}} = \mathbf{z}$.

Following the procedure described in Ref.~\onlinecite{Bookman2015},
the slow evolution of the perturbed droplet's parameters for large NC
radii $\rho_* \gg 1$, weak damping/STT $\sigma = \mathcal{O}(\alpha)
\ll 1$, and low temperature $\beta \ll 1$ is governed by the set of
coupled, stochastic differential equations
\begin{widetext}
  \begin{subequations}
    \label{eqn:fullNLsystem}	
    \begin{align}
      \label{eq:2}
      d{\phi} &=\omega\,dt
      -\frac{\sigma}{4\pi}\int_{\lvert\mathbf{x}\rvert\leq\rho_*}
      (\mathbf{v}\cdot\hat{\boldsymbol{\rho}})
      \, \sech^2\left(\rho-\frac{1}{\omega}\right)
      \hspc[2]d\mathbf{x}dt+d\mathcal{W}_\phi,\\
      d{\boldsymbol{\xi}} &= \mathbf{v}dt +
      \frac{\sigma\omega}{2\pi}\int_{\lvert\mathbf{x}\rvert\leq\rho_*}
      \sech^2\left(\rho-\frac{1}{\omega}\right)\hat{\boldsymbol{\rho}}
      \hspc[2]\hspc[2]
      d\mathbf{x}dt + d\bW_{\bxi},\\
      d\omega & =
      \alpha\omega^2(\omega+h_0)dt-\frac{\sigma\omega^3}{4\pi}
      \int_{\lvert\mathbf{x}\rvert\leq\rho_*}\sech^2
      \left(\rho-\frac{1}{\omega}\right)\hspc[2]
      d\mathbf{x}dt +d\mathcal{W}_\omega,\\
      d{\mathbf{v}} &=
      \alpha\omega\mathbf{v}(\omega+2h_0)dt-
      \frac{\sigma\omega^2}{2\pi}\int_{\lvert\mathbf{x}\rvert\leq\rho_*}
      \left(\frac{3}{2}\mathbf{v}-
        \frac{(\mathbf{v}\cdot\hat{\boldsymbol{\varphi}} 
          )}{\rho\omega} \hat{\boldsymbol{\varphi}}\right)
      \sech^2\left(\rho-\frac{1}{\omega}\right)\hspc[2] d\mathbf{x}dt
      + d\bW_{\bV},
		\end{align}
	\end{subequations}
\end{widetext}
which are to be interpreted in the Stratonovich
sense~\cite{dAquino2006}. There is a symmetry in these equations
between the droplet's collective precessional dynamics and motion.
The phase $\phi$ and position $\boldsymbol{\xi}$ dynamics have a
leading order linear coupling to the frequency $\omega$ and
velocity $\mathbf{v}$ equations, respectively.  The second, additional
terms in the phase and position equations proportional to
  current $\sigma$ correspond to higher order corrections from
soliton perturbation theory, which prove to be essential for
describing perturbed droplet dynamics~\cite{Bookman2013,Bookman2015},
in particular, the finite temperature effects explored here.

The terms ${\mathcal{W}}_i$, with $i=\phi$, $\bxi$, $\omega$, and
$\bV$, are scaled Weiner processes, with nontrivial covariance
structure. Each noise term is a spatial integral of the thermal field
perturbation against an appropriate kernel (see
Ref.~\onlinecite{Bookman2015}, Eqs.~4.1--4.4). If we arrange them into
a vector $\boldsymbol{\mathcal{W}} = \left( \mathcal{W}_\phi \hspc[3]
  \mathcal{W}_{{\xi}_x} \hspace{3 pt}\mathcal{W}_{{\xi}_y} \hspace{3
    pt}\mathcal{W}_\omega \hspace{3 pt}\mathcal{W}_{{v}_x}\hspace{3
    pt} \mathcal{W}_{{v}_y} \right)^T$, then the covariance between
the processes is given by
\begin{equation}
	\text{E}[\boldsymbol{\mathcal{W}}\boldsymbol{\mathcal{W}}^T]
    =
    \frac{\beta^2 t}{2\pi} \left [ \begin{array}{cc}
        \begin{array}{cc} 
          \sigma_\phi^2 & \mathbf{v}^T/2 \\
          \mathbf{v}/2 & \omega \mathbf{I} 
        \end{array}
        & \mathbf{0} \\
        \mathbf{0} &
        \begin{array}{cc}
          \omega^5/2 & \omega^4 \mathbf{v}^T \\
          \omega^4 \mathbf{v} & \boldsymbol{\sigma}_{\mathbf{v}}^2
        \end{array}
    \end{array} \right ],
    % \!=\! 
    % \frac{\beta^2t}{2\pi}
	% \left(
	% \begin{array}{cccccc}
	% 	\sigma^2_\phi & \frac{v_x}{2} & \frac{v_y}{2} & 0 & 0 & 0\\
	% 	\frac{v_x}{2} &  \omega & 0 & 0 & 0 & 0 \\
	% 	\frac{v_y}{2} & 0 & \omega & 0 & 0 & 0  \\
	% 	0 & 0 & 0 & \frac{\omega^5}{2} & \omega^4 v_x & \omega^4v_y \\
	% 	0 & 0 & 0 & \omega^4v_x & \sigma^2_{v,x} &  2\omega^3v_xv_y \\
	% 	0 & 0 & 0 & \omega^4v_y & 2\omega^3v_xv_y & \sigma^2_{v,y}\\
	% \end{array}
	% \right),
	\label{eqn:NLcovariance}
\end{equation}
where we have, for the sake of compactness, denoted
\begin{subequations}
	\begin{align}
      \sigma^2_\phi &= \frac{v^2}{4\omega}+
      \frac{\omega}{2},\\ 
      \boldsymbol{\sigma}^2_{\mathbf{v}} &=
      \left [
        \begin{array}{cc}
          \omega^5+\frac{\omega^3}{4}
      \left(9v_x^2+v_y^2\right) & 2 \omega^3 v_x v_y \\ 
      2 \omega^3 v_x v_y & 
      \omega^5+\frac{\omega^3}{4}\left(v_x^2+9v_y^2\right)
        \end{array}
      \right ]
	\end{align}
\end{subequations}
and $\mathbf{I}$ is the $2\times 2$ identity matrix and $\mathbf{0}$
is the $3 \times 3$ zero matrix.

\section{Deterministic Linearization and Stability}

We will first examine the dynamics of Eqs.~(\ref{eqn:fullNLsystem}) at
zero temperature $\beta^2=0$. These deterministic dynamics have been
studied in detail~\cite{Bookman2013,Bookman2015}. When the current
$\sigma$ exceeds the minimal sustaining current $\sigma_\text{min}$,
the system undergoes a saddle-node bifurcation resulting in a stable
fixed point denoted $({\bxi}_*,\omega_*,{\bV}_*)$ that encapsulates
the balance between damping and STT to sustain the droplet. The stable
fixed point is stationary at the center of the NC,
$\boldsymbol{\xi}_*=\mathbf{v}_*=0$, with precessional frequency
$\omega_*$ determined as a root of the transcendental equation
\beq
\frac{\sigma}{\alpha} = \frac{2(h_0 + \omega_*)}{1 + \omega_*\left[
    \log \left( \frac{1}{2} \textrm{sech}\left( \rho_*
        -\frac{1}{\omega_*} \right) \right) + \rho_* \tanh
    \left(\rho_* - \frac{1}{\omega_*} \right) \right]}. 
\label{eqn:omegastar}
\eeq

\begin{figure}[b]
\centering
\includegraphics[width=0.8\linewidth]{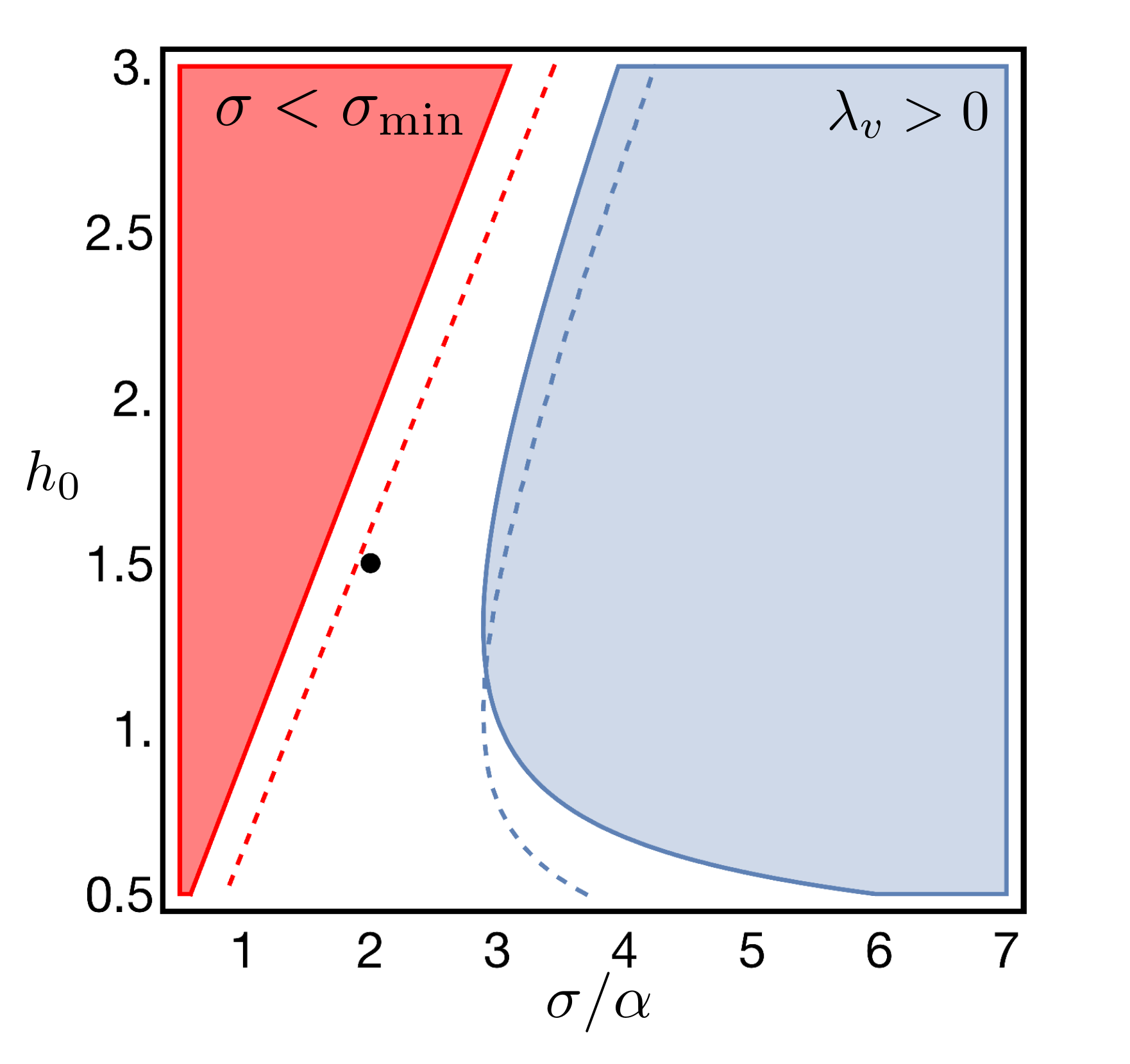}
\caption{\label{fig:stableparams} (color online) Droplet existence and
  linearly stable parameter space for a droplet nucleated in a NC of
  normalized radius $\rho_*=15$. The droplet cannot exist in the left
  region (filled red) where $\sigma<\sigma_\mathrm{min}$, whereas the
  droplet is linearly unstable in the right region (filled blue) where
  $\lambda_v>0$. Therefore, the droplet is stable in the remaining
  white region. The numerical simulations in
  Sec.~\ref{sec:numer-simul} and Figs.~\ref{fig:sample_path},
  \ref{fig:sxi}, \ref{fig:linewidth}(a) with $h_0=1.5$, $\alpha =
  0.03$, and $\sigma=2\alpha$ (black circle) exhibit linear
  stability. The dashed lines are boundaries for droplet existence and
  linear stability with reduced NC radius $\rho_*=5$.}
\end{figure}

We observe that the phase $\phi$ in Eqs.~\eqref{eqn:fullNLsystem}
decouples from the system, so its dynamics can be determined from the
remaining three parameters. If we linearize
Eqs.~(\ref{eqn:fullNLsystem}) around this fixed point, we arrive at
the system
\begin{subequations}
	\begin{align}
	\dot{\phi} &= \omega,\\
	\dot{\boldsymbol{\xi}} &= \mathbf{v} + \lambda_\xi\boldsymbol{\xi},\\
	\dot{\omega} &= \lambda_\omega(\omega-\omega_*),\\
	\dot{\mathbf{v}} &= \lambda_V\mathbf{v},
	\end{align}
\end{subequations}
where 
\begin{subequations}
  \label{eqn:eigenlin}
  \begin{align}
	\lambda_\xi &=
    -\frac{1}{2}\sigma\rho_*\omega_*\sech^2
    \left(\rho_*-\frac{1}{\omega_*}\right),\\  
	\lambda_\omega &= -h_0\alpha\omega_*+\lambda_\xi+ 
    \frac{1}{2}\sigma\omega_*
    \left(\tanh\left(\rho_*-\frac{1}{\omega_*}\right)+1\right),\\
	\lambda_v &= -2\alpha\omega_*^2+\lambda_\omega-\lambda_\xi.
	\end{align}
\end{subequations}
It is necessary to carefully choose parameters so that this fixed
point is stable, \emph{i.e.}, so that all eigenvalues in
Eq.~(\ref{eqn:eigenlin}) are negative. The condition $\sigma>\alpha
h_0$ is sufficient for $\lambda_\xi,\lambda_\omega<0$, but in order to
ensure that $\lambda_v<0$, we require additionally that
\begin{equation}
  \alpha\omega_*(2\omega_*+h_0) >
  \frac{1}{2}\sigma\omega_*
  \left(\tanh\left(\rho_*-\frac{1}{\omega_*}\right)+1\right). 
\label{eqn:stabcond}
\end{equation}
Note that the inequality requirement for stability in
\eqref{eqn:stabcond} was not identified previously~\cite{Bookman2015},
and is essential to understanding the dynamics of the droplet. It is
possible to visualize the region of linear stability in the
($h_0$,$\sigma/\alpha$) plane as in Fig.~\ref{fig:stableparams}. The
left (red) area corresponds to the condition
$\sigma<\sigma_\mathrm{min}$, where the droplet cannot exist. This
approximately linear relation for the existence boundary has been
corroborated by experiment~\cite{Macia2014}.  The inequality
requirement Eq.~(\ref{eqn:stabcond}) adds an unstable, right region (blue
area) where the velocity of the droplet increases until it drifts away
from the NC area and damping destroys it. The remaining white area
represents the parameter space where the droplet exists and is
stable. We observe that such a region shifts to lower applied
fields and increased current for smaller NC radii (dashed lines).

It is helpful to express these eigenvalues in a more tractable form,
so that we can observe how they approximately scale with experimental
parameters. For this, we define a parameter-dependent constant that we
will denote by $\zeta$
\begin{equation}
	\zeta = \frac{2\alpha h_0}{\sigma}-1.
\end{equation}
Previous work~\cite{Bookman2015} assumed that the current was near the
critical value $\sigma \approx 2 \alpha h_0$, so that
$\zeta=\mathcal{O}(\rho_*^{-1})$. This work relaxes that assumption
and allows for any current that is sufficiently above the minimum
sustaining current so that Eq.~\eqref{eqn:omegastar} can be
approximately inverted to obtain the frequency tunability
\begin{equation}
  \label{eq:4}
  \omega_* = \rho_*^{-1} + \mathrm{arctanh} \, (\zeta) \rho_*^{-2} +
  \mathcal{O}(\rho_*^{-3}) . 
\end{equation}
Then the leading-order approximations of each eigenvalue are
\begin{subequations}
  \label{eq:3}
  \begin{align}
    \lambda_\xi &=
    -\frac{\sigma}{2}(1 - \zeta^2)+\mathcal{O}(\rho_*^{-1}),\\ 
    \lambda_\omega &=
    -\frac{\sigma}{2}(1-\zeta^2)+\mathcal{O}(\rho_*^{-1}),\\ 
    \lambda_v &=
    \mathcal{O}(\sigma\rho_*^{-2}). 
	\end{align}
\end{subequations}
The expression for $\lambda_v$ is prohibitively complex, so we omit it
here.

Inequality \eqref{eqn:stabcond} is a fundamental result identifying a
deterministic mechanism that can drive droplet drift instability.  Any
nonzero $\mathbf{v}$ (recall that $\mathbf{v} \ne 0$ corresponds to a
spatial phase gradient across the droplet in Eq.~\eqref{eq:5}) will
slowly increase when Eq.~\eqref{eqn:stabcond} does not hold.
Counterintuitively, large applied current destabilizes the droplet.

Previous work~\cite{Bookman2013} has analyzed the dynamics of this
system with $\mathbf{v}\equiv0$, but we find here that the inclusion
of the velocity dynamics is essential. The dynamics are much more
sensitive to the choice of parameters, as is seen both above in the
linear case, and below in the full nonlinear case. A key observation
is that while $\lambda_\xi,\lambda_\omega$ are small
$\mathcal{O}(\sigma)$, $\lambda_v$ is much smaller $\mathcal{O}(\sigma
\rho_*^{-2})$. These eigenvalues dictate the relaxation rate of the
system towards the fixed point.  When compared to the $\mathbf{v} = 0$
dynamics, the relaxation rate of the droplet center $\boldsymbol{\xi}$
decreases by a factor proportional to $\rho_*^{-2} \ll 1$
when the $\mathbf{v}$ dynamics are included. Furthermore, we see that
$\lambda_v$ can change sign, while $\lambda_\xi$ and $\lambda_\omega$
are negative for $\sigma > \alpha h_0$. This suggests that there
is a shallow basin of attraction for the fixed point, allowing for the
possibility of linear drift instabilities mediated by thermal
noise. Indeed, all experiments~\cite{Mohseni2013,Macia2014,Backes2015}
have been performed outside the region of linear stability, suggesting
droplet drift instability and the concomitant observation of
low-frequency spectral features.  We note that the theory presented
here is nominally applicable to the case $\rho_* \gg 1$, whereas the
experiments in Refs.~\onlinecite{Mohseni2013,Macia2014,Backes2015}
with $\rho_* \in (5,8)$ are at the borderline of applicability.

\section{Stochastic linearization}

One method of approximating the dynamics of the stochastic system
Eqs.~(\ref{eqn:fullNLsystem}) is to employ the previously calculated
linearization of the deterministic system and approximate the noise,
now denoted $\boldsymbol{\mathcal{W^*}}$, by
evaluating the covariance matrix, Eq.~(\ref{eqn:NLcovariance}), at the
fixed point. This low temperature theory yields the linear
stochastic system
\begin{subequations}
	\begin{align}
	d\phi &= \omega dt + d\mathcal{W}_\phi^*,\\
	d{\boldsymbol{\xi}} &= \bV dt+\lambda_\xi\boldsymbol{\xi}dt +
    d\bW_{\bxi}^*,\\ 
	d{\omega} &= \lambda_\omega(\omega-\omega_*)dt +d\mathcal{W}_{\omega}^*,\\
	d{\mathbf{v}} &= \lambda_v \mathbf{v}dt + d\bW_{\bV}^*.
	\end{align}
	\label{eqn:noisylinearsystem}
\end{subequations}
When evaluated at the fixed point, the covariance matrix becomes
diagonal
\begin{align}
  \text{E}[\boldsymbol{\mathcal{W^*}}\boldsymbol{\mathcal{W}^*}^T]&=
  \beta^2t\cdot\text{Diag}\left(\frac{\omega_*}{4\pi},
    \frac{\omega_*}{2\pi},\frac{\omega_*}{2\pi},
    \frac{\omega_*^5}{4\pi},\frac{\omega_*^5}{2\pi},
    \frac{\omega_*^5}{2\pi}\right),\nonumber
  \\ 
  &\equiv t\cdot\text{Diag} \left(\beta_\phi^2,\beta_\xi^2,
    \beta_\xi^2,\beta_\omega^2,\beta_v^2,\beta_v^2\right),
  \label{eqn:lincovariance}
\end{align}
where we denote the variance of each parameter by $\beta_i^2$ for
$i=\phi$, $\bxi$, $\omega$, and $\bV$. The linear system
Eqs.~(\ref{eqn:noisylinearsystem}) can be solved explicitly. For
$(\boldsymbol{\xi},\omega,\mathbf{v})$, we obtain a set of coupled O-U
processes that describe the stochastic properties of the linear
system.  
% It is worth pointing out that the precession amplitude of a
% spatially uniform, finite temperature NC-STO with easy-plane
% anisotropy has also been described by an O-U
% process~\cite{Silva2010}.

Of particular interest is the behavior of the decoupled oscillator
phase, $\phi(t)$, as it allows us to relate our analytical description
with the generation linewidth, $\Delta f$, which can be measured from
the electrical characterization of NC-STOs~\cite{Dumas2014}. Solving
the system in Eqs.~(\ref{eqn:noisylinearsystem}), we find that the
frequency is an O-U process with mean $\omega_*$ and variance
\begin{align}
  \text{Var}[\omega(t)] &= -\frac{\beta_\omega^2}{2\lambda_\omega}
  \left(1-\exp{2\lambda_\omega t}\right)\nonumber\\ 
  &\rightarrow
  -\frac{\beta^2_\omega}{2\lambda_\omega},\quad\text{as}\,\,
  t\rightarrow\infty.  
\end{align}
We can then write down the solution for the phase $\phi(t)$ as the
  sum of an integrated O-U process and a Wiener process (random walk)
\begin{equation}
  \phi(t) = \int_0^t \omega(s)ds + \mathcal{W}^*_\phi,
\end{equation} 
from which we find that the variance of $\phi$ quickly approaches
linear growth
\begin{equation}
  \text{Var}[\phi(t)]\rightarrow
  \left(\frac{\beta_\omega^2}{\lambda_\omega^2}+\beta_\phi^2\right)t
  \quad\text{as}\,\,t\rightarrow\infty, 
\end{equation}
so that the spectral lineshape is Lorentzian and the generation
linewidth is given by
\begin{align}
  \Delta f &=
  \left(\frac{\beta_\omega^2}{\lambda_\omega^2}+\beta_\phi^2\right) =
  {\beta^2}\left(
    \frac{\omega_*^5}{4\pi\lambda_\omega^2}+\frac{\omega_*}{2\pi}\right). 
  \label{eqn:linearlinewidth}
\end{align}

By virtue of Eq.~(\ref{eqn:beta2}), the generation linewidth is
linearly proportional to temperature. Because $\omega_*\ll 1$, the
generation linewidth is dominated by the Wiener process, phase noise
contribution [second term in Eq.~(\ref{eqn:linearlinewidth})]
resulting from the higher order contribution to the phase dynamics of
Eq.~\eqref{eq:2}. This expression for generation linewidth is also
consistent with the notion of a reduced impact of thermal fluctuations
on a larger magnetic mode volume. Indeed, recalling Eq.~\eqref{eq:4},
it is clear that larger NC radii minimize the generation
linewidth.
% Similarly, the phase of an easy-plane, finite temperature, spatially
% uniform NC-STO undergoes a random walk~\cite{Silva2010}.

We are also interested in the dynamics of the center
$\boldsymbol{\xi}$ as it describes the droplet's random motion with
respect to the NC region. The velocity and position form a coupled
pair of O-U processes, which we can solve using standard methods. We
then find the variance of the droplet center
\begin{align}
  s_{\bxi}^2(t) &=-\frac{\beta _{\xi }^2}{2 \lambda _{\xi }}
  \left(1-e^{2 \lambda _{\xi }t}\right)-\frac{\beta _v^2}{2
    \left(\lambda _{\xi }-\lambda _v\right){}^2}\nonumber\\ 
  &\quad \times\left(\frac{1-e^{2 \lambda _{\xi }t}}{\lambda _{\xi
      }}+\frac{4 \left(1-e^{\left(\lambda _{\xi }+\lambda
            _v\right)t}\right)}{\lambda _{\xi }+\lambda
      _v}+\frac{1-e^{2 \lambda _v t}}{\lambda _v}\right)\nonumber\\ 
  &\rightarrow 
  -\frac{1}{2}\frac{\beta_v^2}{\lambda_\xi^2\lambda_v}-
      \frac{\beta_\xi^2}{\lambda_\xi} 
  \quad \text{as}\,\,t\rightarrow\infty. 
  \label{eqn:lineardrift}
\end{align}
We might expect that the position noise term in
Eq.~(\ref{eqn:lineardrift}) would dominate, analogous to the phase
noise in Eq.~(\ref{eqn:linearlinewidth}). However, the balance of the
two terms in Eq.~(\ref{eqn:lineardrift}) is highly sensitive to
experimental parameters. In fact, for the parameters used in this
study, the velocity noise term is the dominant contribution.

Equations~(\ref{eqn:linearlinewidth}) and (\ref{eqn:lineardrift}) are
central results of this paper. The former relates our stochastic
theory to an experimental observable, namely, the Lorentzian
generation linewidth. The latter quantifies the amount of droplet
drift with respect to the NC center and thus provides a means to
quantify the drift instability from random fluctuations in the
magnetic system.

\section{Numerical Simulations}
\label{sec:numer-simul}

\begin{figure}[t]
  \begin{center}
    \includegraphics{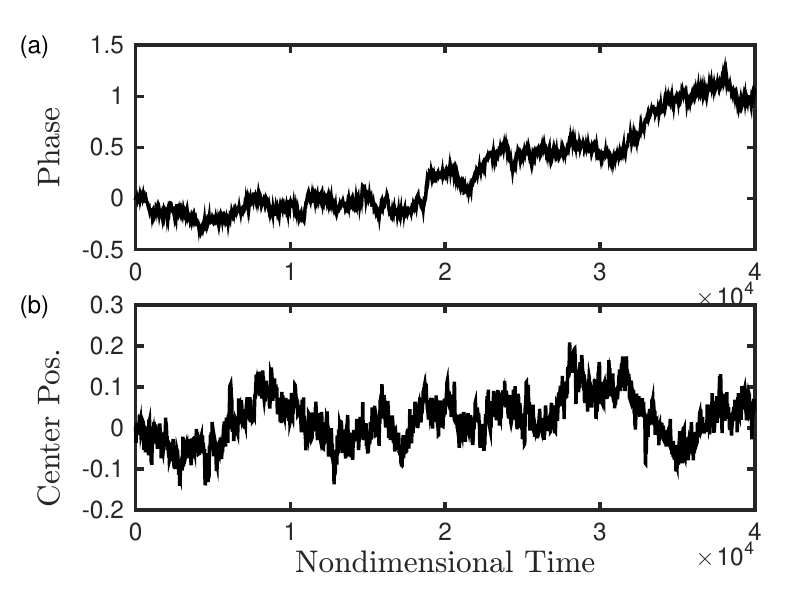}
    \caption{\label{fig:sample_path} Numerically computed nonlinear
      sample path from Eqs.~\eqref{eqn:fullNLsystem} for (a) the
      droplet phase $\phi$ and (b) the $x$-component of the center
      position $\xi_x$. The phase is measured in radians, and position
      and time are nondimensional as per Table~\ref{table:scalings}.}
	\end{center}
\end{figure}

To examine the behavior of the full nonlinear system,
Eqs.~(\ref{eqn:fullNLsystem}), we numerically simulate an ensemble of
sample paths. Details of our numerical implementation can be found in
the Appendix. We choose the parameters $\rho_* = 15$, $h_0=1.5$,
$\alpha = 0.03$, and $\sigma=2\alpha$ in order to ensure that we are
within the asymptotic validity of our analysis and the region of
linear stability, depicted by the black dot in
Fig.~\ref{fig:stableparams}. Typical sample paths of the droplet's
phase and center position generated by this method are shown in
Fig.~\ref{fig:sample_path}.

\begin{figure}[t]
  \begin{center}
    \includegraphics{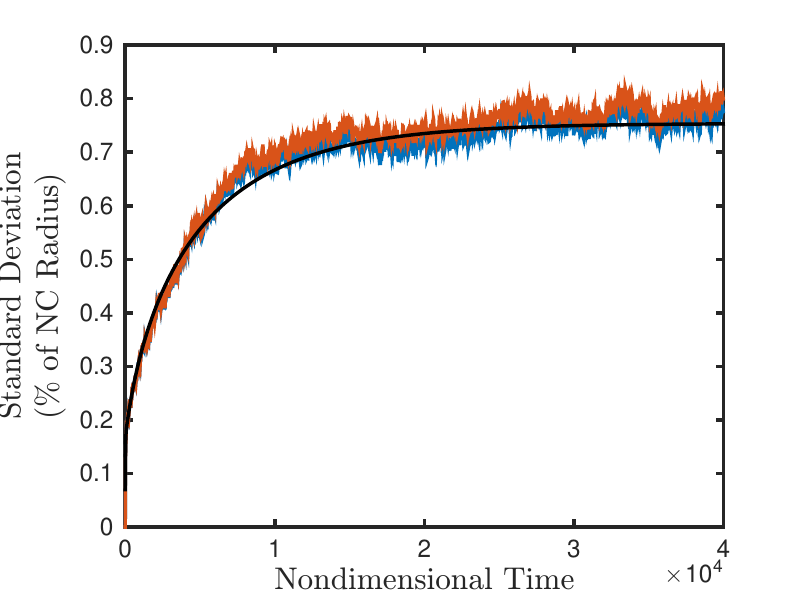}
	\caption{\label{fig:sxi} (color online) Standard deviation
      $s_{\bxi}$ of the droplet center from linear theory (solid black
      line), linear simulation (solid blue), and nonlinear simulation
      (solid red). }
  \end{center}
\end{figure}

We first examine the statistics of the droplet center.
Figure~\ref{fig:sxi} shows the standard deviation of the droplet
center for an ensemble of numerical simulations of the linear (blue)
and nonlinear (red) systems. The analytical prediction of
Eq.~(\ref{eqn:lineardrift}) (black) agrees well with the linear
simulation. For the chosen set of parameters, nonlinearity is not
observed to significantly enhance the droplet drift and, in fact, the
standard deviation of the droplet center from the NC center is never
more than 1\% of the NC radius.  For slightly modified parameters, we
observe qualitatively different behavior when nonlinearity is
introduced. For example, reducing the NC radius to $\rho_*=10$, we
approach the stability boundary of Fig.~\ref{fig:stableparams},
although linear stability in Eq.~(\ref{eqn:stabcond}) is still
satisfied. However, the numerical simulation of the nonlinear system
leads to approximately 15\% of the simulated paths leaving the NC
before $t=4\cdot10^{4}$. This indicates that the basin of attraction
of the system is relatively small. We can also infer from the
simulations at larger NC radii that the size of the basin of
attraction decreases with NC radius. This suggests that the small NC
devices used in experiments at room temperature sustain droplets that
exhibit deterministic or thermally induced drift instabilities during
measurements. In fact, typical spectral
measurements~\cite{Mohseni2013,Lendinez2015} acquire data in time
spans on the order of seconds, which translate to $\approx 1\cdot
10^{10}$ in our normalized units. The characterization of the
multi-dimensional boundary in phase space of the basin of attraction
and ejection statistics are, however, outside the scope of this
paper. Note that in the ensemble used to generate Fig.~\ref{fig:sxi},
no sample paths ejected from the NC.

\begin{figure}[b]
  \centering
  \includegraphics[width=3.3in]{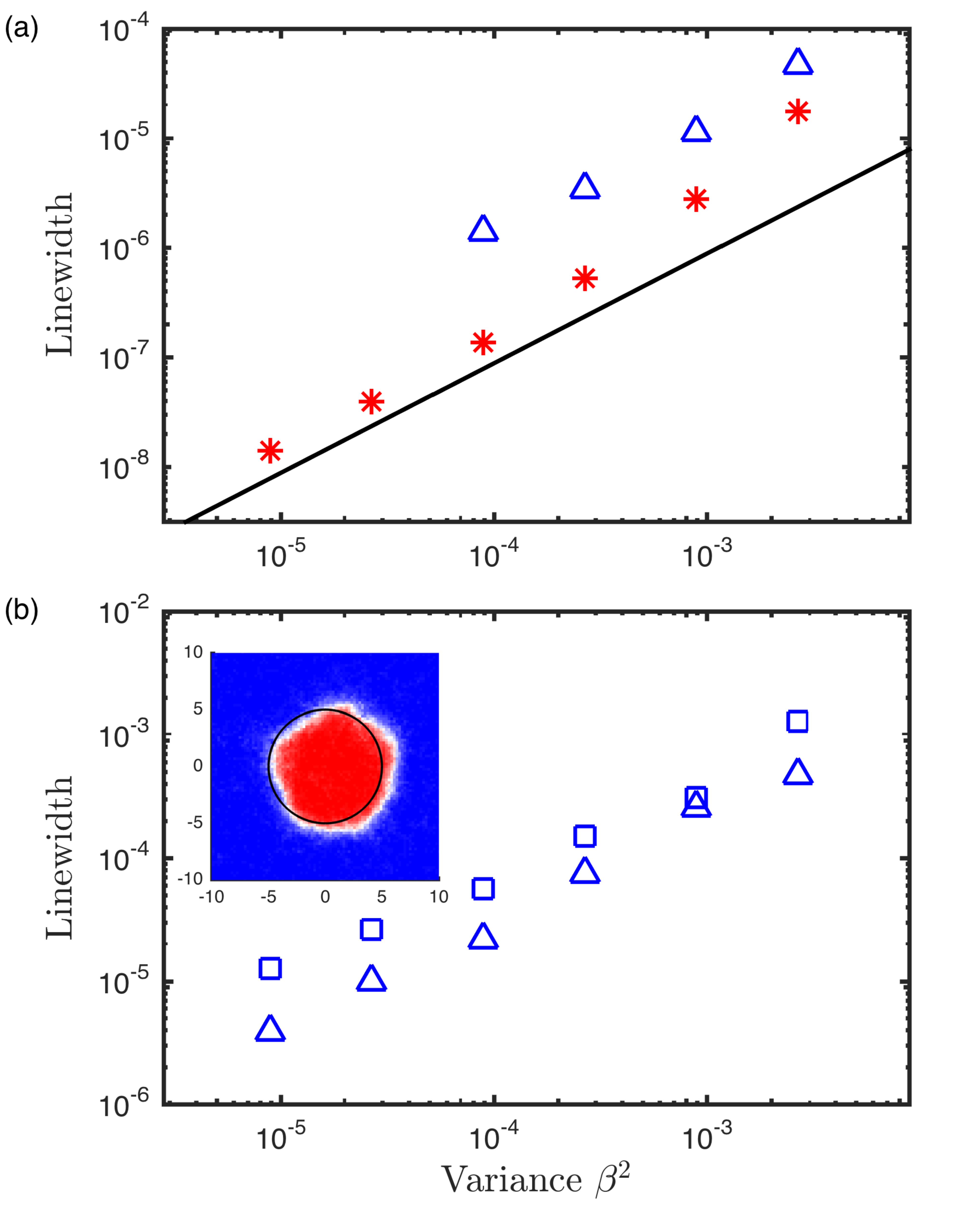}
  \caption{\label{fig:linewidth} (color online) (a) Droplet linewidth
    as a function of temperature from linear theory (solid black
    line), nonlinear simulation (red asterisks), and micromagnetic
    simulations (blue triangles) when $\rho_*=15$, $h_0 = 1.5$,
    $\alpha = 0.03$, $\sigma = 2\alpha$. (b) Droplet linewidth as a
    function of temperature from micromagnetic simulations where only
    the NC radius is reduced to $\rho_*=5$ from (a) (blue triangles)
    and the effect of a current-induced Oersted field (blue
    squares). Linewidth is expressed in rad/$\tau$ as per
    Table~\ref{table:scalings}. Inset shows droplet profile with
    Oersted field included. Error bars are $\mathcal{O}(10^{-9})$ and
    are not shown.}
\end{figure}

In the regime where the droplet does not drift away from the NC over
the timescale simulated, it is possible to compare the linear
generation linewidth to numerical simulations of
Eqs.~(\ref{eqn:fullNLsystem}). From a sample path of the stochastic
phase $\phi(t)$, we calculate the linewidth via the power spectral
density of $\phi$, as discussed in Ref.~\onlinecite{Silva2010}. It is
worth noting that the linewidth calculated via this method is strictly
valid for white noise~\cite{Eklund2014} and can vary between sample
paths due to the fluctuations between each path. For the linewidths
reported here, we take the mean value of the calculated linewidths
from $500$ different sample paths. Figure~\ref{fig:linewidth}(a) shows
the linewidth's dependence on temperature for the nonlinear system
(red asterisks) and Eq.~(\ref{eqn:linearlinewidth}) (solid black
line). Finite sampling and the asymmetric, heavy-tailed distribution of linewidths across sample paths causes the mean to converge slowly to the linear theory at low temperature. Although the median gives results more clearly convergent to the linear theory, the mean corresponds to experimentally observed linewidths, which are averaged over long timescales. Nonlinear simulations yield a mean linewidth of $1.77\cdot10^{-5}$ at
temperature $\beta^2 = 2.8\cdot10^{-3}$, which, for comparative
purposes, corresponds to 214 kHz at temperature $T = 314$ K under the
temporal and temperature scalings of Ref.~\onlinecite{Mohseni2013}
with damping enhanced to $\alpha=0.03$.  Note, however, that the
material parameters ($\rho_*$, $\sigma$, and $h_0$) for the nonlinear
simulations do not correspond to those from
Ref.~\onlinecite{Mohseni2013}.

The linear theory is a very good predictor of the nonlinear system's
behavior at low temperatures and we numerically observe that the
discrepancy between the linear and nonlinear linewidths decreases
quadratically in $T$ as $T\rightarrow 0$, as one would expect from
this perturbative approach. However, as room temperature is
approached, the nonlinear linewidth exceeds the linear linewidth by an
order of magnitude. This originates from the increased impact of
thermal fluctuations when the linearization is not strictly
applicable. We stress that current experiments have not directly detected
ejection events, and in the event of ejection, the bias current can
re-nucleate a droplet and the resulting linewidth in aggregate will be
considerably broader due to the ensuing transient dynamics. Our
simulations end upon ejection, and do not allow for re-nucleation.

The above simulations are strictly valid for the regime $\rho_* \gg 1$
with negligible long-range dipole and Oersted fields. Experiments to
date, however, have been performed when $\rho_* \in (5,8)$.  Moreover,
it is important to characterize the impact of dipolar and Oersted
fields on the droplet's collective motion and precession. To further
explore droplet behavior, we perform full-scale micromagnetic
simulations with non-local dipole fields using the GPU-based package
Mumax3~\cite{Vansteenkiste2014}. We first compare micromagnetic
results by choosing the same set of dimensionless parameters specified
above and scalings consistent with Co/Ni
multilayers~\cite{Mohseni2013} (See Table~\ref{table:scalings}).
% , namely, saturation magnetization $M_s = 756$~kA/m, exchange
% stiffness $A=13$~pJ/m, and perpendicular magnetic anisotropy
% $K_u=519.3$~J/m$^3$.
The fixed layer is assumed to be perpendicularly polarized.
% Dimensionalizing the chosen set of parameters, we obtain a NC radius
% of $133$~nm, and applied flux of $0.6$~T, and a driving current
% density of $1.3\times10^{13}$~A/m$^2$.
%The NC is placed at the geometrical center of a $832\times 832\times
%3.6$~nm active area discretized in cells with size $3.25\times
%3.25\times 3.6$~nm, below the exchange length.  
The NC is placed at the geometrical center of an active area of size
$89.9\times 89.9\times 0.39$ discretized in cells with size $0.35\times
0.35\times 0.39$, below the exchange length. An ensemble of sample paths
is not feasible to compute micromagnetically due to time constraints,
so we determine the linewidth from a single path spanning $t =
1.8\times 10^4$ and sampled at $0.015$ intervals.
% $1.5$~$\mu$s and sampled at $1.25$~ps.
First, we do not observe droplet motion, which is consistent with the
results shown in Fig.~\ref{fig:sxi} where the droplet center variance
is expected to be below our cell resolution. The results for the
temperature dependent linewidth are shown in
Fig.~\ref{fig:linewidth}(a) as blue triangles. We note that the
micromagnetic simulations overestimate the linewidth obtained from
nonlinear simulations but are on the same order of magnitude at room
temperatures. At low temperatures, the micromagnetic simulations do
not approach the linear theory as one would expect.  This is a
consequence of the limited simulation time and the spatial resolution of
our micromagnetic scheme %, and the deterioration of numerical micromagnetic schemes in general for stochastic problems~\cite{martinez_minimizing_2007}
that precludes an accurate
estimation of the phase noise statistics and thus its convergence to
the linear linewidth.

Despite this limitation, micromagnetic simulations can be used to
explore the dynamics of droplets sustained in devices with smaller NC
radii, where micromagnetics have shown to be more
accurate~\cite{Mohseni2013,Lendinez2015} and where, conversely, the
theory is not strictly applicable. We perform micromagnetic
simulations with the same nondimensional parameters specified above
but reduce the NC radius to $\rho_*=5$, in the range of experiments
performed to date, and increase the current to $\sigma=0.1$. The
resulting linewidths are shown in Fig.~\ref{fig:linewidth}(b) by blue
triangles. A qualitative agreement with theory is observed, namely, a
linear dependence of the linewidth on temperature and a linewidth
increase for smaller NC radii.  Additionally, micromagnetic
simulations allow us to include the current-generated Oersted
field~\cite{Petit2012b,Dumas2013,Madami2015}. This further enhances
the linewidth by a factor $\sim 5$ (blue squares) originating from
the distortion of the droplet boundary, as observed from a snapshot of
the $\mathbf{\hat{z}}$ magnetization component shown in the inset of
Fig.~\ref{fig:linewidth}(b). These results suggests that the unavoidable non-local
and Oersted fields in a real device will enhance the generation
linewidth compared to theory, but the temperature-dependent features
remain mostly unchanged.

\section{Discussion and Conclusion}

We have developed a stochastic perturbation theory for magnetic
dissipative droplets describing the random motion of the droplet's
position, velocity, frequency, and phase.  Higher-order perturbative
effects in the phase and position are shown to be essential for
understanding the dynamics of the droplet.  Inclusion of velocity
dynamics causes a qualitative shift in the behavior of the droplet
position, and gives rise to a previously uncovered deterministic drift
instability. Such an instability occurs at high driving currents,
leading to an exponential increase in the droplet velocity. This
counter-intuitive effect also implies a small basin of attraction for
the stable fixed point, providing a simple explanation for the origin
of drift instabilities from randomness in the system, such as thermal
fluctuations.

We find that in parameter regimes where the deterministic droplet is
linearly stable, the stochastically induced drift instabilities are
rare events compared with the typical precessional timescales. A
notable implication is that the observation of drift instabilities due
to thermal fluctuations using micromagnetic simulations is
prohibitive.  In contrast, our finite dimensional reduction of the
governing partial differential equation makes such effects
computationally feasible. The study of rare events is beyond the scope
of this paper, but motivates an application of large deviation theory,
as previously studied, for example, in the context of fiber optic
soliton communication systems~\cite{Moore2008}. Likewise,
micromagnetic simulations tailored to study rare
events~\cite{Vogler2013} might be used to resolve the time and
computational limitations. Even in the deterministic case, the
predicted linear instability may be difficult to recover from
micromagnetic simulations due its slow rate of exponential growth.
From an experimental point of view, typical measurement timescales
suggest that drift instability and droplet renucleation can occur many
times. For example, the long timescale required in the direct imaging
of localized excitations, 500 ms, indicates that drift instabilities
could occur $\sim 10^6$ times, leading to the small droplet amplitude
and spatial smearing observed in the XMCD images of
Ref.~\onlinecite{Backes2015}.

In contrast, previous works have interpreted the droplet drift
mechanism through spatial inhomogeneities in field~\cite{Bookman2015}
or anisotropy~\cite{Lendinez2015}.  Here, we have identified two
additional drift mechanisms, a deterministic linear instability
inherent to the NC-STO system and rare drift events caused by thermal
fluctuations.

Our model also allows us to obtain an analytical expression for the
linearly stable droplet generation linewidth.  At low temperature, we
find that the phase noise is characterized by a Wiener process (random
walk) and the droplet center is an O-U process, analogous to the
stochastic phase and amplitude dynamics, respectively, of spatially
uniform STOs~\cite{Silva2010}.  For the linearized system, the
resulting generation linewidth is linearly dependent on temperature,
whereas the nonlinear system exhibits a linewidth enhancement when
approaching room temperature, reflecting the coupling between the
droplet's constituent variables. Full-scale micromagnetic simulation,
including the fully nonlinear spatial variation of the system,
qualitatively agree with the numerical results. However, we do not
observe convergence toward the linear theory at low temperatures using
a standard micromagnetic package~\cite{Vansteenkiste2014}. This
suggests the study of droplet generation linewidth as a test problem
for stochastic micromagnetic codes~\cite{ragusa_full_2009}.

The analytical and numerical linewidths obtained are two orders of
magnitude below the typical linewidths observed in experiments. This
disagreement may be caused by the small NC radii used experimentally,
the existence of non-local dipolar and current-induced Oersted fields,
and the aforementioned drift instabilities for data-acquisition
timescales. In fact, micromagnetic simulations performed with a radius
similar to those experimentally fabricated to date return linewidths
in the same order of magnitude when both non-local and current-induced
Oersted fields are included. The relevance of such fields in the
generation linewidth motivates their inclusion in the analytical theory.
For thin films, the effect of non-local dipole fields on deterministic
droplet dynamics has been shown to be a frequency downshift when
$\mathbf{v} = 0$~\cite{Bookman2013}.  It remains to incorporate these
effects into the stochastic theory when $\mathbf{v} \ne 0$.  Because
the Oersted field is not a singular perturbation~\cite{Bookman2015},
its inclusion in this collective theory would necessitate the
incorporation of droplet coupling to spin waves.  Such coupling is in
principle possible, see, e.g., Ref.~\onlinecite{kath_soliton_1995}.

In conclusion, this work provides the means to seek optimized
experimental parameters for a given application. To wit, we find that
an environment with a large NC radius, low field, modest current, and
large anisotropy are less susceptible to drift and thus lead to a much
narrower generation linewidth. Our results motivate a more detailed
experimental study on the current and temperature-dependent generation
linewidth and ejection statistics of droplets.

\begin{acknowledgments}
	E.~I. acknowledges support from the Swedish Research Council,
    Reg. No. 637-2014-6863.  M.~A.~H. was partially supported by NSF CAREER
    DMS-1255422.
\end{acknowledgments}

\bibliographystyle{aipnum4-1}
%\bibliography{thermaldroplet2015}
%

\section*{Appendix}
\appendix
\section{Numerical Methods}

We simulate the nonlinear system Eq.~(\ref{eqn:fullNLsystem}) via the
Euler-Maruyama method, with drift correction to account for the
Stratonovich interpretation of the stochastic
integrals~\cite{kloeden1992}. We use a timestep of $dt = 4$, and our
total integration time is $t = 4\cdot10^4$. We integrate 500
sample paths, and then use the standard sample variance to
  produce Figure~\ref{fig:sxi}.

\begin{figure}[b]
	\centering
	\includegraphics{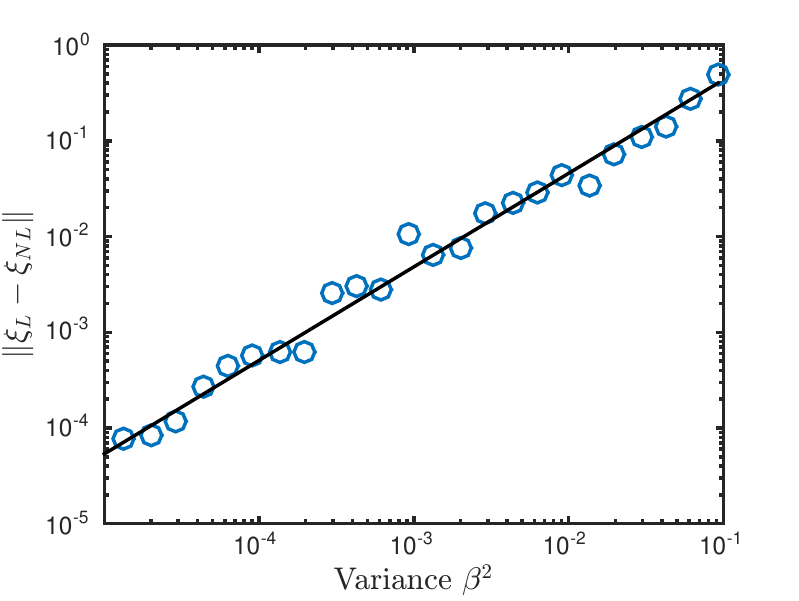}
	\caption{(color online) Convergence plot indicating that
		$|\xi_\text{NL}-\xi_\text{L}|=\mathcal{O}(\beta^2) =
		\mathcal{O}(T)$. Best-fit line has slope of 0.9793.}
	\label{fig:convergence}
\end{figure}

We must ensure that our nonlinear and linear systems coincide when
$T\rightarrow0$. To that end, we calculate the pathwise difference
between the droplet center $\boldsymbol{\xi}_\text{L}$ calculated by
discretizing the linear system Eq.~(\ref{eqn:noisylinearsystem}) and the
droplet center $\boldsymbol{\xi}_\text{NL}$ calculated via
discretizing the nonlinear system Eq.~(\ref{eqn:fullNLsystem}). Note that
both paths are calculated using the same stochastic terms, scaled
appropriately. The results are shown in
Figure~\ref{fig:convergence}. The standard deviation of the droplet
center from the fixed point is $\mathcal{O}(\sqrt{T})$, and the
separation between the nonlinear and linear paths is $\mathcal{O}(T)$,
so we have
\begin{equation}
  \|\boldsymbol{\xi}_\text{NL}(t)-\boldsymbol{\xi}_\text{L}(t)\| =
  \mathcal{O}(s ^2_{\xi}). 
\end{equation}
This linear convergence in $T$ is a positive consistency check on the
linearization Eq.~\eqref{eqn:noisylinearsystem} and stochastic
timestepping of the nonlinear system Eq.~\eqref{eqn:fullNLsystem}.

\end{document}